\documentclass[12pt]{iopart}

\usepackage{graphicx}
\usepackage{lineno}
\begin{document}

\title{H-theorem bypassing by 2D Brownian particle under discontinuous magnetic field}

\author{P M Romanets}

\vspace{10pt}
\begin{indented}
\item[]April 2022
\end{indented}

\begin{abstract}
 In the present paper, we theoretically study the kinetic properties of 2D charged particles under a discontinuous magnetic field. It is shown that certain conditions could cause their bypassing of the H-theorem. We use the classical kinetic equation to derive the diffusion equation for the case. The essential postulates are made for modeling: (i) the motion of the particles is restricted by a rectangular region; (ii) the mobility of the particles goes to zero while they came closer to borders; (iii) the Coulomb interaction between particles is absent; (iv)the interaction between charged particles and the medium doesn’t change the total momentum of the medium. The achievability of the conditions is discussed.

\end{abstract}

%
%
%
%
%

\section{Introduction}
The breakdown of H-theorem could make a revolution in world energy and environmental problems. Also, it removes many restrictions for the technological progress of modern storage devices. For information theory, the H-theorem means that storage devices tend to lose a part of the information with time. In general, only the breakdown of the H-theorem can kill this harmful effect. Therefore, the problem of great actuality, and equally great hopelessness.

One of the earliest attempts to break down the H-theorem was the Brownian ratchet. Afterward, Marian Smoluchowski noted that it won't work because thermal fluctuations also act on the parts of the machine \cite{Smoluchowski}. There are one more fundamentals why it fails: it is the third Newton's law. An opposite force act on all the particles that give the net amount of work. In other words, the total angular momentum of the molecules and mechanism remains constant. Therefore, the probability to get the net amount of work from fluctuation will decrease with time while it is not zero.

The very remarkable experimental work of A. Schilling et al. partially breaks the statement about heat flow direction \cite{ExperimentsBasedOnPeltier}.  But the general statement about entropy gain is not broken. Indeed,  one should include both the heat transfer part and the electrical part to calculate full entropy gain.

The breakdown of the H-theorem is impossible. But one can mimic it by weakening the conditions of the theorem. Namely, the system is not closed but interacts with one or a few other large systems (reservoir). We must emphasize, from a quantum point of view, the H-theorem statement implies entanglement between the system quantum state and the reservoir quantum state (entanglement also could be the hence of measurements, other way the entropy alweys be constant \cite{Neuman,QThermodynamic, MaxwellDemon}).   But the energy and momentum exchange between them should be negligibly small. Additional ways to bypass the H-theorem arise when we consider the system under an external magnetic field.  G.B. Lesovik et al. considered a quantum spin-polarized system and showed a gain of the entropy by the magnetic field for this case \cite{H-theorem_in_qp}. Invetigation performed by George S. Levy within classical approach but in the same time no less interesting. The cobmination of elstic scattering by surface and external magnetic filed were leading to the cycloidal motion of charged particles. The authors focused their investigation on the Onsager relations and CPT symmetry.
\begin{figure}
	\centering
	\includegraphics[width=80mm]{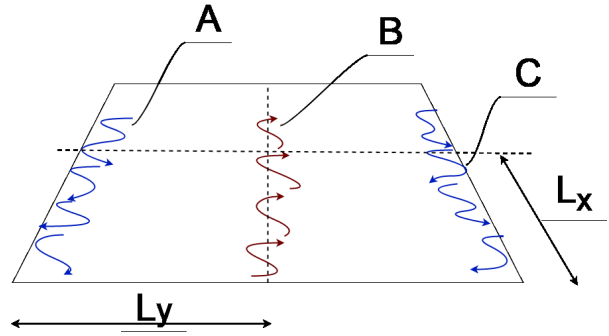}
	\caption{The scheme shows the trajectories arising due to the magnetic field and surface scattering (A, C) and the trajectories arising due to the magnetic field step near $y=0$ (B).}
	
\end{figure}
In the present study, we move further and consider 2D charged particles interacting with the medium and subjected to a strongly inhomogeneous magnetic field with a step-like profile.  We will suppose that interaction between particle and medium corresponds to the diffusion regime. We also neglect any Coulomb interaction between charged particles therefore they behave like Brownian particles. 

\section{Diffusion equation}
 The Boltzmann equation separated into symmetric and asymmetric parts relative to the velocity has the next form:
\begin{eqnarray}
\frac{\partial f_s}{\partial t} + [\omega_c \times {\bf v}]\frac{\partial f_s}{\partial {\bf v}} + {\bf v}\frac{\partial f_a}{\partial {\bf r}} = 0;\nonumber\\
\frac{\partial f_a}{\partial t} + [\omega_c \times {\bf v}]\frac{\partial f_a}{\partial {\bf v}} + {\bf v}\frac{\partial f_s}{\partial {\bf r}} = -\nu f_a.\label{eq1}
\end{eqnarray}
The cyclotron frequancy $\omega_c=\omega_{c0}sgn_{\delta}(y)$ (see eq.(\ref{eq3})).  The distribution functions $f_a(t,{\bf r}, {\bf v}) = -f_a(t,{\bf r}, -{\bf v})$ and  $f_s(t,{\bf r}, {\bf v}) = f_s(t,{\bf r}, -{\bf v})$. There is no energy relaxation integral because we suppose that the particle is already thermalized. The solution of this equation will describe both the cyclotron oscillation on low time scales $t\le\nu^{-1}$ and the slow evolution on time scales $t\gg\nu^{-1}$. Our goal is to consider the second one. We restric our consideration with diffusion regime. The approach is correct when the number of collisions between particles of medium and charged particles per characteristic time of evolution is large. More rigorously, one should apply limitations on distribution function as done below. Supposing the smallness of the $|f_a|\ll f_s$ and the slowness of the $f_s$ ($|\partial f_s/\partial t| \ll \nu |f_a|$) it is easy to derive:
\begin{eqnarray}
\frac{\partial f_s}{\partial t} = \frac{\partial }{\partial x}\left(D_l\frac{\partial f_s}{\partial x}+D_t\frac{\partial f_s}{\partial y}\right) + \frac{\partial }{\partial y}\left(D_l\frac{\partial f_s}{\partial y}-D_t\frac{\partial f_s}{\partial x}\right),\label{eq2}
\end{eqnarray}
where $D_l=T~m^{-1}\nu/(\nu^2+\omega_c^2)$ and $D_t=T~m^{-1}\omega_c/(\nu^2+\omega_c^2)$ are the coordinate dependent diffusion coeficients. The diffusion coeficients are dependent on the temperature $T$ and the mass $m$ of the particle.
We will solve this equation numerically inside the rectangular region $-L_x\leq x\leq L_x$ and $-L_y\leq y\leq L_y$. 
But before, we proceed with the qualitative discussion of the equation aimed to narrow an interesting phenomena search.
First of all, we are interested in how to obtain non-homogeneous distribution in a stationary regime $t\gg L_{x,y}^2 / |D_{l,t}|$.One can see that there is always a trivial solution when the initial distribution is homogeneous. To understand why one could imagine the diffusion process as the scattering between different ballistic trajectories (see Fig.1). The snake-like trajectory (B) is due to the magnetic field step. The two other trajectories are due to the magnetic field and surface scattering. In the case of the homogeneous distribution, the currents compensate each other ${\bf j}_A+{\bf j}_B+{\bf j}_C=0$. We will choose an initial distribution that goes to zero near the region edges to avoid this family of solutions. In addition, we will suppose charged particles' mobility goes to zero near the region edges. Thus we will obtain ${\bf j}_{A, C}=0$. One more thing must be noted: the fenomenon we are studying connected to the magnetic feild step (line $y=0$). While charged particle does not cross this step we will observe the usual Brownian motion slowed by a magnetic field. Therefore we are not considering narrow initial distribution when usual diffusion takes an essential part of the evolution.
Summarizing all the above, we put:
\begin{eqnarray}
f_s(t=0,{\bf r}) = {\cal N} sgn_{\gamma}\left(L_y-|y|\right)sgn_{\gamma}\left(L_x-|x|\right),\nonumber\\\nu({\bf r}) = \frac{\nu_0}{sgn_{\beta}\left(L_y-|y|\right)sgn_{\beta}\left(L_x-|x|\right)},\nonumber\\
\omega_c(y) = \omega_{c0}sgn_{\delta}\left(y\right), \label{eq3}
\end{eqnarray}
where ${\cal N}$  is the normalization constant, $sgn_{\alpha}(z)=\tanh(z/\alpha)$, and $\beta$, $\gamma$, $\delta$, $\nu_0$ and $\omega_{c0}$ are numerical parameters. The numerical calculation is performed using the finite difference method. For the calculation, we use units of time $\omega_{c0}^{-1}$ and  length $\omega_{c0}^{-1}\sqrt{T/m}$. The problem also could be scaled as the standard diffusion problem, $t\to a t$ and ${\bf r}\to b{\bf r}$ where $a=b^2$. The distribution is normalized to one particle. The entropy is calculated according to the Boltzmann formula $S(t) = -\int_S f_s\ln(f_s) d^2{\bf r}$ (we omit the Boltzmann constant for convenience).  The results of the calculation are presented in Fig.2. Figure 2(a,b) shows the average x-position of the particle vs time, i.e.  $<x(t)>=\int_S x f_s d^2{\bf r}$. The frame (a) corresponds to zero mobility on the edge. One can see that the effect of the magnetic field step increases with the increase of the ratios $\omega_{c0}/\nu_0$ and $L_x/L_y$. The increase of the first one means increasing the mobility or the magnetic field strength. The second one refers to the geometry of the considered region. Obviously, the geometrical probability for the particle to be scattered between snake-like trajectories (denoted as B in Fig.1) is higher for the higher ratio $L_x/L_y$. In other words, particles are more likely to cross the line $y=0$ during evolution when the ratio $L_x/L_y$ is large. The frame (b) corresponds to non-zero mobility on the edge, when $\nu=\nu_0$. One can see that some feeble effects take place in this case only before diffusion make the distribution homogeneous. In this case, stationary distribution is always homogeneous. The frame (c) in  Fig.2 shows the gain of the entropy. It is clearly seen that the entropy gain is negative. The frame (d) in Fig.2 shows a countur plot for the stationary distribution function. This is the case of the square geometry of the region. The distribution is heart-shaped. Its bottom sharp part is because of snakes-like trajectories and its rounded upper parts are because of usual diffusion processes.
\begin{figure}
	\centering
	\includegraphics[width=80mm]{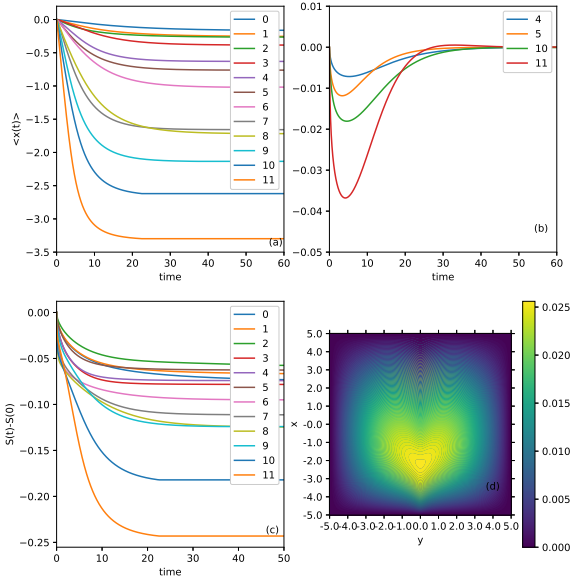}
	\caption{Frames (a): averaged x-position for the particle vs time, the case of eqs. (\ref{eq3}); (b): the same for the case when $\nu=\nu_0$, and  ${\bf j}_{A, C}\neq0$; (c): the gain of the entropy, the case of eqs. (\ref{eq3});  (d): the countur plot for the stationar distribution $f_s(t\to\infty,{\bf r})$ (the parameters are the same as for the line 4). We use $\gamma=0.25$, $\beta=0.25$ and $\delta=0.2$ for (\ref{eq3}). The parameters $L_{x,y}$ and $\nu_0$ are difineted by the lines numbering as next: 
		(0): $\nu_0=2$, $L_x=5$, $L_y=10$
		(1): $\nu_0=1$, $L_x=5$, $L_y=10$
		(2): $\nu_0=0.5$, $L_x=5$, $L_y=10$
		(3): $\nu_0=2$, $L_x=5$,$L_y=5$
		(4): $\nu_0=1$, $L_x=5$,$L_y=5$
		(5): $\nu_0=0.5$, $L_x=5$,$Ly=5$
		(6): $\nu_0=2$, $L_x=5$,$L_y=2$
		(7): $\nu_0=1$, $L_x=5$,$L_y=2$
		(8): $\nu_0=2$, $L_x=5$,$L_y=1$
		(9): $\nu_0=0.5$, $L_x=5$,$L_y=2$
		(10): $\nu_0=1$, $L_x=5$,$L_y=1$
		(11): $\nu_0=0.5$, $L_x=5$,$L_y=1$  .
	
 }
	
\end{figure}

\section{Conclution}
We have considered the diffusion of the charged particles under a discontinuous magnetic field. The results are supposed to be applicable to the 2D electron gas in semiconductor quantum well but not restricted to. We have not taken into account the effects of electron gas degeneration. Corresponding effects will decrease diffusion coefficients. It could be taken into account by the corresponding scaling.

It has been demonstrated that the entropy gain for the charged particle is negative. Nevertheless, the phenomenon could not be considered the breakdown of the H-theorem because the system of charged particles is not enclosed. Because of third Newton's law (or momentum conservation law), the total momentum of the charged particles and the particles of the medium remains zero during the evolution. Similarly, as it takes place for the Brownian ratchet. But this harmful effect should be considered on a case-by-case basis. For example, in semiconductors, it is related to the phonon drag effect, where the effective temperature gradient is $3T<x>/L_x^2$ \cite{PhononDrag1,PhononDrag2, PhononDrag3, PhononDrag4}. Thus one could expect that after evolution time  $>L_x^2/(<x>s_l)$ the phonon drag effect could quench the considered phenomenon, where $s_l$ is the longitudinal acoustic phonon velocity.

\section*{Acknowledgement}
{\it I want to thank all those who support the Ukrainian people in this terrible war. Thank all the heroic soldiers who are defending our country. I express my sincere condolences to all those whose loved ones died.}

\end{document}